# The orbital motion, absolute mass, and high-altitude winds of exoplanet HD209458b

Ignas A. G. Snellen[1], Remco J. de Kok[2], Ernst J. W. de Mooij[1], Simon Albrecht[3,1]

[1]*Leiden Observatory, Leiden University, Postbus 9513, 2300 RA Leiden, NL*

[2]*SRON, Sorbonnelaan 2, 3584 CA Utrecht, The Netherlands*

[3]*Department of Physics, and Kavli Institute for Astrophysics and Space Research, Massachusetts Institute of Technology, Cambridge, Massachusetts 02139, USA*

**For extrasolar planets discovered using the radial velocity method[1], the spectral characterization of the host star leads to a mass-estimate of the star and subsequently of the orbiting planet. In contrast, if also the orbital velocity of the planet would be known, the masses of both star and planet could be determined directly using Newton's law of gravity, just as in the case of stellar double-line eclipsing binaries. Here we report on the detection of the orbital velocity of extrasolar planet HD209458b. High dispersion ground-based spectroscopy during a transit of this planet reveals absorption lines from carbon monoxide produced in the planet atmosphere, which shift significantly in wavelength due to the change in the radial component of the planet orbital velocity. These observations result in a mass determination of the star and planet of $1.00 \pm 0.22$ $M_{sun}$ and $0.64 \pm 0.09$ $M_{jup}$ respectively. A ~ 2 km sec$^{-1}$ blueshift of the carbon monoxide signal with respect to the systemic velocity of the host star suggests the presence of a strong wind flowing from the irradiated dayside to the non-irradiated nightside of the planet within the 0.01-0.1 mbar atmospheric pressure range probed by these observations. The strength of the carbon monoxide signal suggests a CO mixing ratio of $1\text{-}3 \times 10^{-3}$ in this planet's upper atmosphere.**

We observed HD209458 (V=7.65) for ~5 hours on the night of August 7, 2009, using the Very Large Telescope of the European Southern Observatory at Cerro Paranal in Chile. The observations covered the 180 min. transit plus a 40 and 90 min. baseline before and after the event. We used the cryogenic high-resolution infrared echelle spectrograph CRIRES[2], located at the Nasmyth A focus of UT1, making use of the Multi-Application Curvature Adaptive Optics system MACAO[3].

We obtained 51 spectra with a wavelength coverage of 2291 to 2349 nm at a spectral resolution of 100,000. For a detailed description of the observational set-up and data reduction we refer the reader to the supplementary information. During the transit, star-light filters through the atmosphere of the planet, leaving an imprint of molecular absorption lines in the spectrum. In the observed wavelength regime, 56 strong spectral lines from carbon monoxide are expected to be present, and we extracted the CO-signal by cross-correlating a CO model-spectrum with the observed data. We developed our own transmission model to calculate the expected CO-spectrum, with the planet atmosphere described by one mean profile that is in hydrostatic equilibrium (see the supplementary information for details). In our initial model, the temperature is based on the best-fit day-side temperature profile[4] for HD209458b, and gases are uniformly mixed with volume mixing-ratios of $2 \times 10^{-4}$ ($CH_4$ and CO) and $5 \times 10^{-4}$ ($H_2O$) as based on the case[5,6] of HD189733b.

The observed spectra are completely dominated by numerous telluric absorption lines, caused mainly by methane and water vapour in the Earth's atmosphere[7]. The depths of these lines vary with airmass, and an important part of the data reduction process therefore involves the removal of this telluric contamination. Residual effects, clearly present at the positions of strong telluric lines, are further suppressed by normalizing each pixel value in a spectrum by its variance over time. Although this may in certain places also suppress carbon monoxide lines, it prevents the cross-correlation signal to be dominated by telluric residuals. Our analysis results in the significant detection of a CO signal and detection of the orbital motion of the planet. Similar cross-correlation analyses with $H_2O$ and $CH_4$ templates did not result in detections. The model transmission spectrum of $CH_4$ is so densely packed with lines that it strongly hampers a cross-correlation analysis. Two $CH_4$ lines are significantly stronger than the others, but unfortunately one falls in a gap between two of the detector arrays. Cross-correlating with a model spectrum combining CO with $H_2O$ and $CH_4$ does not give an improvement over the signal from CO alone.

The result of the CO cross-correlation is shown in Figure 1. Due to the orbital velocity of the planet, of which the radial component changes during the transit, the CO-signal shifts in position from the beginning to the end of the transit by ~30 km sec$^{-1}$. As shown in Figure 2, from this shift we derive the planet orbital velocity to be $v_p = 140 \pm 10$ km sec$^{-1}$ (1σ), corresponding to the maximum radial velocity at quadrature. In combination with the velocity of the host star due to its orbit around the centre of mass of the system, masses of both star and planet can now be directly solved for using solely Newton's law of gravitation, just as in the case of stellar double-line eclipsing binaries. Large data sets of radial velocity measurements are available for the star[8] showing a sinusoidal variation with an amplitude[9] of 84.3±1.0 m/s, consistent with a zero eccentricity, which is also constrained by the timing of the planet's secondary eclipse[10]. Assuming an orbital period of P=3.5247 days and a planet orbital inclination of 86.93º ± 0.01º (1σ), which are well determined by transit observations[11], the mass of the star is determined to be $M_1 = 1.00 \pm 0.22$ $M_{sun}$, the mass of the planet to be $M_2 = 0.64 \pm 0.09$ $M_{jup}$ at an orbital distance of $a = 0.045 \pm 0.003$ AU (all 1σ uncertainties). It is encouraging to know that the different estimates of the mass of the host star from spectral modelling, such as[12] $M_1$=1.14 ± 0.10 $M_{sun}$, and[13] $M_1$=1.06 ± 0.10 $M_{sun}$, are fully consistent with this.

Several detections of molecular signatures in hot Jupiter atmospheres, from water[14-18], methane[5,18], carbon monoxide[16] and carbon dioxide[18], have been presented in recent literature. Based on observations with the Hubble Space Telescope and/or Spitzer Space Telescope, they lacked the spectral resolution to detect the molecular lines directly as presented here, but instead they targeted the broadband absorption from the rotational-vibrational transition bands. Our CO signal, integrated over the transit assuming a planet orbital velocity of 140 km sec$^{-1}$, is shown in Figure 3. It has a significance of 5.6σ. This means that on average per spectrum the CO signal as presented in Figure 1 and 2 is present at a 1σ level.

Our direct detection of absorption lines leads to an unambiguous identification of carbon monoxide in the atmosphere of HD209458b, and allows us to determine the CO abundance. Although transmission spectra are significantly less dependent on the thermal structure of the atmosphere than dayside spectra, the uncertainty in the CO volume mixing ratio is dominated by the uncertainty in the planet's pressure-temperature profile, and by the uncertainty in the level of masking of CO by $CH_4$ in the atmosphere. The amplitude of the cross-correlation signal is a factor 2.8 stronger than the signal expected from our initial transmission model (with a CO volume mixing ratio of $2\times10^{-4}$). This was determined by adding the initial model CO-spectrum with varying multiplication factors to our data early on in the reduction process, and by subsequently measuring and comparing the strength of the resulting cross-correlation signals. We produced models with varying CO, $CH_4$, and $H_2O$ mixing ratios and atmospheric temperatures, and compared the amplitude of their cross-correlation signals with that observed. This results in a CO volume mixing ratio of $1-3\times10^{-3}$. In the same way we converted the non-detections of $H_2O$ and $CH_4$ to $3\sigma$ upper limits for the water and methane volume mixing ratios of $3\times10^{-3}$ and $8\times10^{-4}$ respectively. Abundance estimates from a dayside spectrum[18] suggest that the mixing ratios of $CH_4$, $H_2O$ and $CO_2$ ($2-20\times10^{-5}$, $0.1-10\times10^{-5}$, and $0.1-1\times10^{-5}$ respectively) are all significantly lower than what we derive for CO. Although the derived abundances are preliminary, our models suggest that the C/H ratio in the upper atmosphere of HD209458b is a factor 2-6 higher than that of the parent star.

Gaussian fitting of the integrated CO-signal shows that it appears blue-shifted with respect to the systemic velocity of the host star by ~2 km sec$^{-1}$, indicative of atmospheric dynamics. To assess the significance of the blue-shift we added series of model CO-spectra to the data as above, but with random systemic velocities for the host star between ±50 km sec$^{-1}$. We determined the $1\sigma$ uncertainty of the radial velocity of the CO-signal to be 1 km sec$^{-1}$ by comparing the offsets between the injected and measured velocities. This indicates that the observed blue-shift of the CO-signal is statistically significant at a 95% confidence level. Sound speeds in the upper layers of hot Jupiters[19,20] are expected to be typically 3-4 km sec$^{-1}$, and winds at a substantial fraction of this speed are indeed possible. Since with transmission spectroscopy we probe the atmospheric region near the planet's terminator, the blue-shift indicates a velocity-flow from the day side to the night side at pressures in the range 0.01-0.1 mbar as probed by these observations. Such winds may be driven by the large incident heat flux from the star on the dayside. Indeed three-dimensional circulation models[20] indicate that at low pressure (<10 mbar) air should flow from the substellar point towards the antistellar point both along the equator and the poles.

**Acknowledgements** We thank the ESO support staff of the Paranal Observatory for their help during the observations. Based on observations collected at the European Southern Observatory (383.C-0045A). S.A. acknowledges support by a Rubicon fellowship from the Netherlands Organisation for Scientific Research (NWO).

**Author Contributions** IAGS participated in the development of the concept of this research and the analysis code, participated in the observations, the analysis and interpretation of the data and writing the manuscript. RdK developed the planet atmosphere models and participated in the analysis and interpretation of the data and writing the manuscript. EdM participated in the development of the concept of this research, and participated in the analysis and interpretation of the data and writing the manuscript. SA participated in the development of the concept of this research, and participated in the analysis and interpretation of the data and writing the manuscript.

**Author Information** Reprints and permissions information is available at www.nature.com/reprints. The authors declare no competing financial interests. Correspondence and requests for materials should be addressed to I.S. (e-mail: snellen@strw.leidenuniv.nl).

Figure 1. **Carbon monoxide signal in the transmission spectrum of exoplanet HD209458b.** The cross-correlation is shown between a template spectrum of 56 CO lines and VLT spectra of HD209458 taken between a planet orbital phase of -0.025 < θ < 0.035. The beginning and end of the transit are at θ ± 0.018. The systemic velocity[21] of the host star HD209458a is -14.77 km sec$^{-1}$ (blue-shifted), and the velocity of the Paranal observatory in the direction of the star is 11.0 km sec$^{-1}$ at the time of observation. This means that a planet CO-signal is expected to be blue-shifted by ~26 km sec$^{-1}$ at mid-transit, exactly where a faint signal is present in the data. Panel **a** and **b** show the same data, with the linear grey-scales indicating the cross-correlation signal (dark means absorption). In panel **a** the cross-correlation as function of the geocentric radial velocity is plotted, and in panel **b** in the rest-frame of the host star showing the CO-signal in the centre. During the transit, the planet signal moves by 30 km sec$^{-1}$ due to the change in the radial component of the planet orbital velocity. For the cross-correlation in panel **c**, our initial model transmission spectrum of CO was added to the data at 3 times the nominal level, to demonstrate the resemblance with the observed signal.

Figure 2. **The expected carbon monoxide signal as function of the planet orbital velocity.** The observed CO signal is shown in grey-scales as in Figure 1. The dotted lines indicate the expected change in radial velocity of the planet over the transit for orbital velocities of 50, 100 and 150 km sec$^{-1}$. We determined the planet orbital velocity to be 140±10 km sec-1 (1σ) using chi-squared analysis. Both the orbital velocities of the planet and star around the planet-star center-of-mass are known, allowing the masses of both objects to be determined to be $m_1$=1.00±0.22 $M_{sun}$ and $m_2$=0.64±0.09 $M_{jup}$ (1σ) using solely Newton's law of gravitation.

Figure 3. **The carbon monoxide signal integrated over the transit**. The cross-correlation signal from all spectra taken during the transit were combined assuming a planet orbital velocity of 140 km sec$^{-1}$, individually weighted by the depth of the transit signal at the observed epoch. The integrated signal is statistically significant at a 5.6σ confidence level. We derive a CO volume mixing ratio of 1-3x10$^{-3}$ for the upper atmosphere of HD209458b, with the precision governed by the uncertainty in the pressure-temperature profile and in the level of masking of the CO signal by $CH_4$. The CO-signal is blue-shifted by ~2 km sec$^{-1}$ with respect to the systemic velocity of the host star, which suggests a velocity-flow from the day side to the night side driven by the large incident heat flux on the day side.

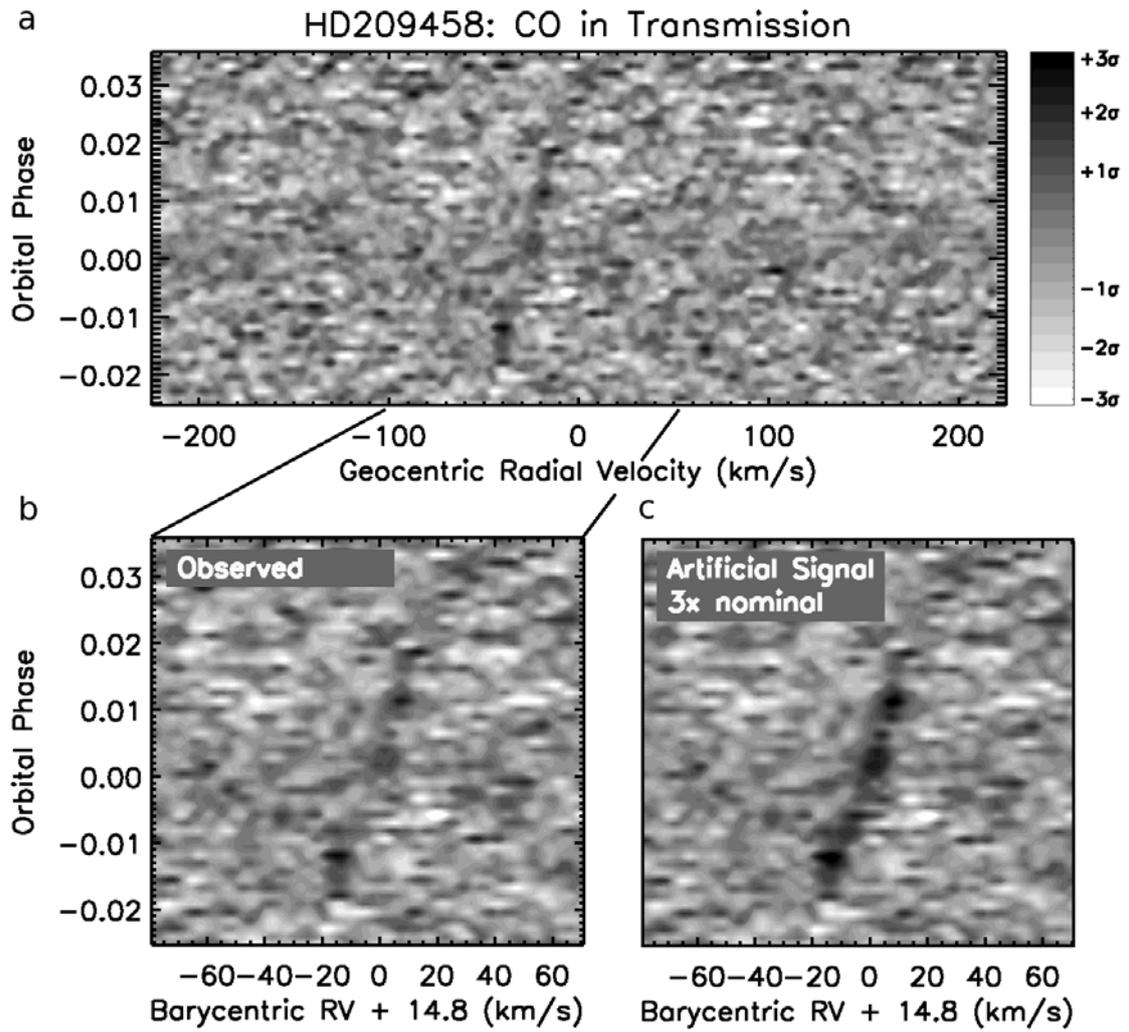

**Fig.1**

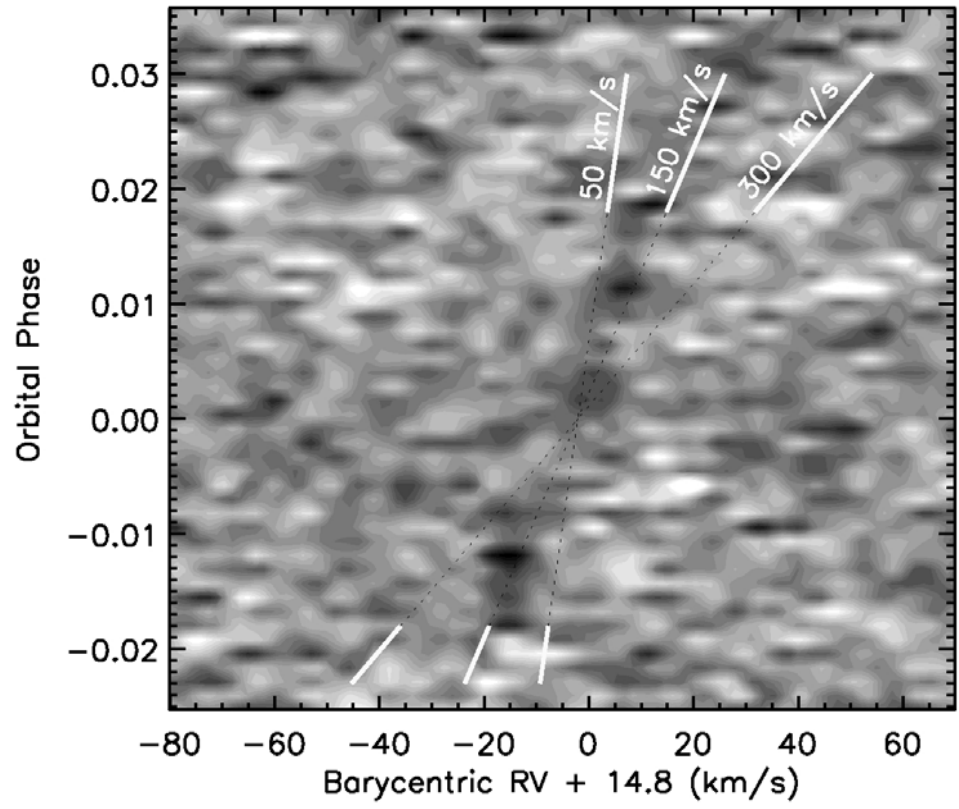

**Fig.2**

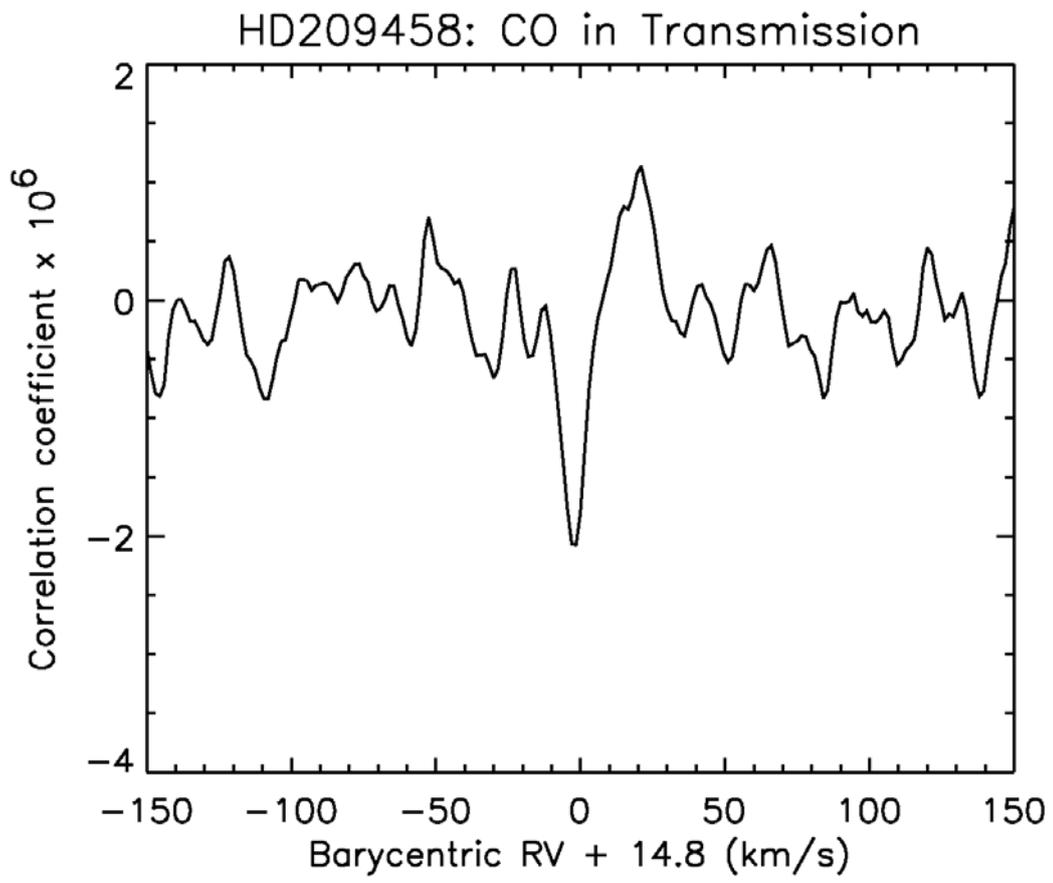

**Fig.3**

## Supplementary information

Ignas A. G. Snellen, Remco J. de Kok, Ernst J. W. de Mooij, Simon Albrecht

**Observational set-up**

We observed HD209458 (V=7.65) for ~5 hours on the night of August 7, 2009, using the Very Large Telescope of the European Southern Observatory at Cerro Paranal in Chile. The observations consist of 51 spectra covering the 180 min. transit plus a 40 and 90 min. baseline before and after the event. We used the CRyogenic high-resolution Infrared Echelle Spectrograph CRIRES[2], located at the Nasmyth A focus of UT1, making use of the Multi-Application Curvature Adaptive Optics system MACAO[3]. We utilized a setup with a central wavelength of 2322 nm, with the spectra imaged on a detector mosaic of four Aladdin III detectors (4x1024x512 pixel) with a gap of 250 pixels between the chips. It provided a wavelength coverage of 2291 to 2349 nm, at a spectral resolution of 100,000 using a 0.2" slit. Each individual spectrum consists of two sets of 5x30 sec exposures with the target nodded back and forwards by 10 arcseconds along the slit.

**Data reduction and analysis: removal of instrumental and telluric effects**

The data were reduced using the CRIRES pipeline[22] V1.11.0, which performed a first order wavelength calibration, background subtraction and flatfield correction, and which combined the individual exposures in 51 extracted spectra. Further data analysis was carried out in IDL. We first interpolated over evidently bad pixels, which affected 0.5% of the detectors. The resulting spectra are completely dominated by telluric absorption lines. The position of these telluric lines were first fitted and compared to the list from the HITRAN database[7] to determine an accurate wavelength solution for each spectrum. The spectra were subsequently interpolated to the wavelength solution of the reference spectrum, which was chosen as one of the spectra with the highest signal-to-noise ratio, resulting in small <0.2 pixel shifts. A second order polynomial was fitted to the ratio of each spectrum over the reference spectrum, which was used to get the continuum of all spectra at the same level as that of the reference spectrum. The resulting spectra are shown as "Step 1" in the upper panel of Figure S1. The normalized count-level in a pixel now varies mainly as function of airmass over time, which we removed using a linear fit with airmass for each pixel independently. Those pixels that are expected to be within 3 pixels from the peak of a planet CO line (as determined from the model transmission spectrum – see below), were masked during this fitting process, but the resulting solutions were applied to them. A total of 6% of all pixels were masked in this way. Due to the expected change in the radial component of the orbital velocity of the planet over the transit, pixels from typically only 10 out of 51 spectra were masked out if they were in wavelength near CO in the rest-frame of the host-star. If no masking would have been applied at all, part of the CO signal (about 25% for both the natural and artificial signals) would have been removed by the applied corrections. The resulting spectra are shown as "Step 2" in the second panel of Figure S1.

Although the night was as good as photometric, absorption by the Earth's atmosphere was found not to be a perfect function of the geometric airmass, which are visible as correlated residuals in the second panel of Figure S1, most clearly seen at the position of strong telluric lines. We found these residuals to be different for water and methane lines, the two dominant telluric molecular species in the observed wavelength-range. We removed these remaining residuals by fitting the variation in each pixel to the residuals determined in a strong water and a strong methane line. Subsequently, remaining low-frequency variations were removed by fitting a second order polynomial to each spectrum. In addition, each pixel value was divided by its standard deviation over time. This last step reduces the influence of noisy parts of the spectra (e.g. in the centres of deep telluric lines) on the cross-correlation signal. The resulting data is shown as "Step 3" in the third panel of Figure S1.

**The model transmission spectra**

We developed our own transmission model to calculate the CO model spectrum. Transmission through a spherical atmosphere was calculated for layers between 5 and $1 \times 10^{-6}$ bar, and was integrated over the entire limb of the planet. Below 5 bar the planet was assumed optically thick, which is confirmed by our calculations. The atmosphere at each point is described by one mean profile that is in hydrostatic equilibrium. In our initial model, the temperature is based on the best-fit day-side temperature profile[4] of HD209458b. Gases are uniformly mixed with volume mixing ratios of $2\times10^{-4}$ ($CH_4$ and CO) and $5\times10^{-4}$ ($H_2O$), based on the case[5,6] of HD189733b. Gas opacity is calculated line-by-line using a Voigt line profile. Opacity data for CO and $H_2O$ are obtained from the HITEMP database[23] and that for $CH_4$ come from[7] HITRAN 2008. Although these databases may not contain some of the weakest lines that can contribute to the continuum at high temperatures, they do contain the strongest lines which determine the sensitivity in our analysis. To model the observed wavelength region we include 2,225 CO lines, 11,631 $H_2O$ lines and 15,146 $CH_4$ lines. $H_2$-$H_2$ collision-induced absorption (CIA) is also included[24]. The initial CO model spectrum is made by subtracting the calculated transit transmission spectrum for the case with CO and CIA from the spectrum that includes only CIA. We also calculate templates from spectra that, besides CIA, also include $CH_4$ or $H_2O$, to emulate the cases where these gases mask part of the CO lines. The same procedure was followed for the other gases. All model spectra are shown in Figure S2, highlighting the regions which overlap with the 4 detector arrays.

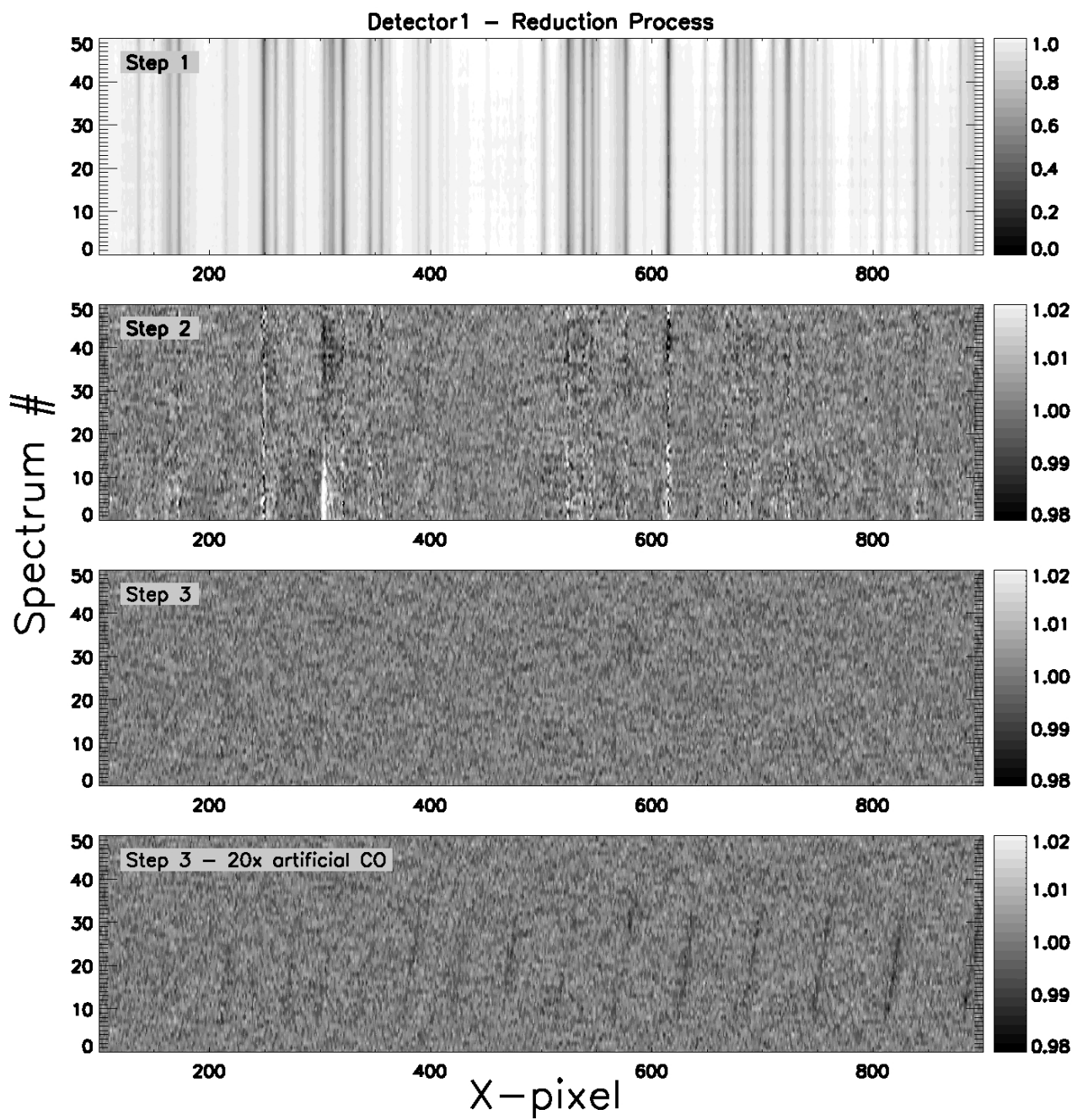

*Fig. S1: The 51 spectra collected with detector 1 showed at different stages in the reduction process. The first panel shows the resulting spectra after they have been normalized and put on the same wavelength scale. The second panel shows the spectra after the dependence on airmass has been taken out and the telluric lines removed. The third panel shows the spectra after the correlated residual effects have been taken out. The individual CO lines are not visible. Since the overall correlated signal from 56 strong lines (combined over 4 detectors) is detected at a 5.6σ level, the individual lines are expected to be present at a significance of 0.7σ, and per spectrum at <0.2σ. The lowest panel shows the same, but with an artificial CO spectrum added to the data at 20x the initial level, hence at 10x the observed level. This clearly reveals the positions of the individual lines.*

**The cross-correlation analysis**

The initial CO model spectrum shows variations in transit depth of ~0.1% over the observed wavelength range. Since our observations and analysis are only sensitive to narrow spectral features (any broad-band feature will have been washed out), we identified 56 CO lines with transit depth >0.004% relative to the local continuum. The strongest of these lines have a transit depth of ~0.04% (see Figure S2). The wavelengths and relative strength of these 56 CO lines, convolved to the resolution of the observed spectra, were used to construct a template spectrum. The 51 residual spectra, as reduced above, consist each of 4 pieces (from the 4 detector-arrays). Each piece was individually cross-correlated with this template spectrum over a velocity range of ±225 km sec$^{-1}$, with time steps of 1.5 km sec$^{-1}$. For each spectrum, the resulting cross-correlation signal consists of the combination of the signals from the four detector arrays, each weight-averaged according to the relative strength of the CO-signal in the template spectrum.

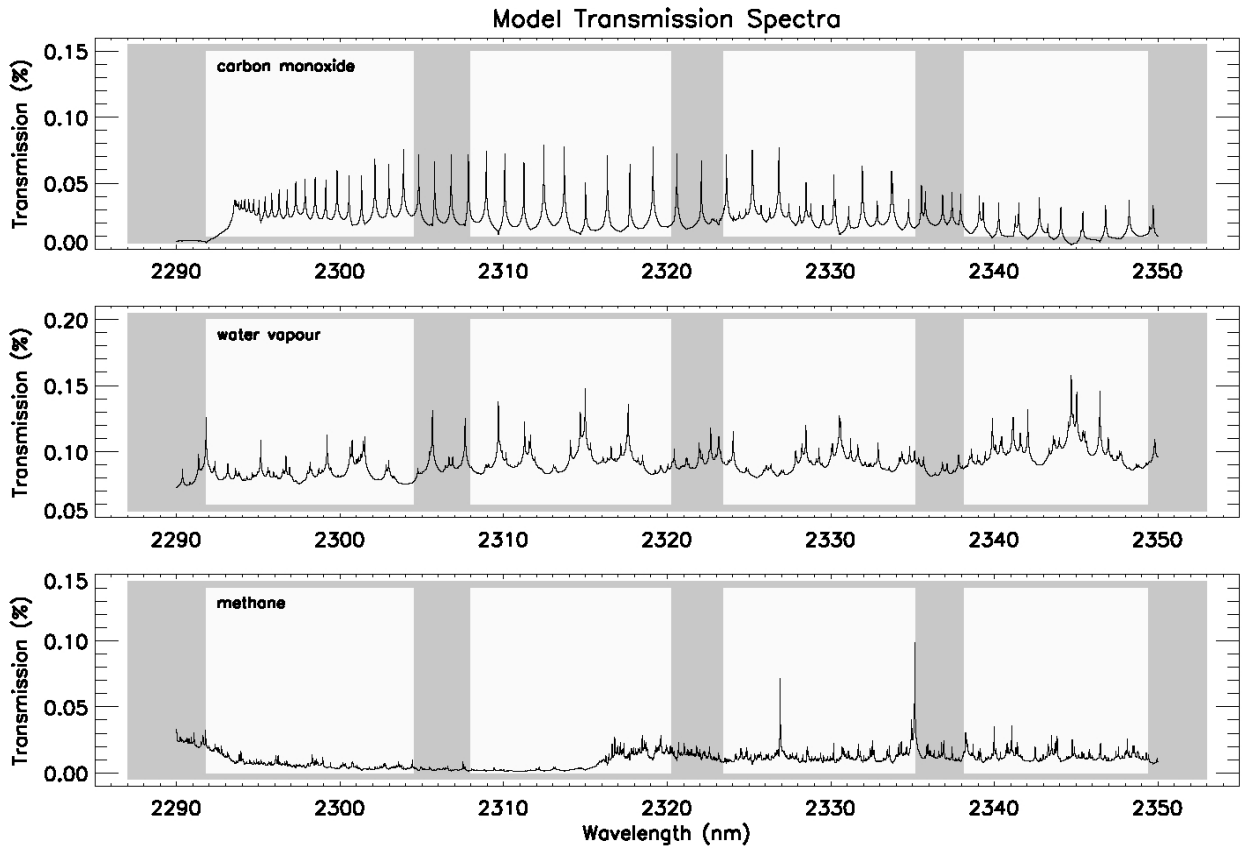

*Figure S2: Models used for the transmission of carbon monoxide (top panel), water vapour (middle panel), and methane (lower panel) in the atmosphere of HD209458b.*

The resulting cross-correlation signal is shown in Figures 1 and 2 with linear grey scales indicating the amplitude (dark corresponding to absorption). The CO signal integrated over the transit is detected at a 5.6σ level, indicating that on average per spectrum the signal is present at a ~1σ level, although this varies significantly from spectrum to spectrum. The latter is caused by the fact that each pixel value was divided by its standard deviation over time. This has as a consequence that over the course of the transit, CO lines move across regions with higher and lower standard deviations (e.g. due to strong telluric features), making the apparent CO signal stronger and weaker over the course of the transit. This is the reason why the strength of the cross-correlation signal can only be converted to physical line-strengths by comparing it to an artificially injected signal added to the observed signal. As one can see in the lower-right panel of Figure 1, the artificial signal brightens and weakens at the same positions in the transit as the observed signal.